\documentclass[showpacs,preprintnumbers,amsmath,amssymb]{revtex4}
\usepackage{graphicx}
\usepackage{dcolumn}
\usepackage{bm}
\begin{document}
\preprint{APS/123-QED}
\title{Comparison of Two Interpretations of Josephson Effect}
\author{I.M.Yurin}
 \email{yurinoffice@mail.ru}
\affiliation{%
I.M.Yurin, Fl.61, bld. 7, 22 Festivalnaya St, Moscow, 125581, Russian Federation}%

\date{\today}

\begin{abstract}
This paper puts forward an interpretation of the Josephson effect
based on the Alternative Theory of Superconductivity (ATS). A
comparison of ATS- and BCS-based interpretations is provided. It is
demonstrated that the ATS-based interpretation, unlike that based on
BCS theory, does not require a revision of fundamentals of quantum
physics.
\end{abstract}

\maketitle

\section{\label{sec:level1}Introduction}

    The Josephson effect holds a special place in theoretical physics.
It could hardly be denied that it was the prediction \cite{1} of
this effect, as well as its subsequent interpretation \cite{2} and
observation \cite{3,4}, that made a stunning impression on the
contemporary academic community, leaving it with no further doubts
as to the validity of the BCS theory.

    Reasons to question the correctness of that theory have not arisen till much later, when
high-temperature superconductors were discovered. Up to that moment,
it was believed based on BCS theory's estimates, that maximum values
of transition temperatures should lie in the vicinity of 40 K.
Although some attempts to suggest alternative versions of
explanation of superconductivity phenomena have been made late in
the last century \cite{5,6,7,8,9}, these attempts were received by
most theorists with skepticism. That is why currently, an
overwhelming majority of works that are taken seriously are related
to BCS theory, and the existing discussions are mostly carried out
at the level of determination of appearance mechanisms of strong
electron-electron attraction.

It would be interesting to trace the connections between the BCS
theory and other concepts by taking the example of the BCS-BEC
crossover \cite{10}, which, according to the authors, should take
place when Cooper pair binding energy values are comparable with
Fermi energies of electrons. It should be noted that the idea of
describing the superconductivity phenomenon using the possibility of
Bose-Einstein condensation (BEC) of electron pairs appeared as such
before Cooper \cite{11,12}, but the linear relationship between the
zero-temperature energy gap $\Delta$ and the transition temperature
$T_c$ of then-known superconductors was difficult to interpret in
the framework of that concept. Therefore, the present-day revival of
this idea is naturally related to the BCS theory.

Curiously enough, the said linear relationship can be explained in
the framework of the Alternative Theory of Superconductivity (ATS),
unrelated to the BCS theory \cite{13}. One should then note that
calculations of transition temperatures in the mean value
approximation are only of illustrative nature in physics, and not
expected to provide high-precision predictions \cite{14}. From this
point of view, if the two theories were to be compared, the relation
$T_c  \approx 2\Delta /3.5$ obtained from the BCS theory does not
provide any advantages as compared with the ATS relation $T_c
\approx \Delta /2$ \cite{13}.

As for the calculation of transition temperature values carried out
in the framework of the ATS \cite{15}, it has unexpectedly revealed
a clear advantage over the BCS theory. Indeed, both theories only
use one adjustable parameter. However, in the BCS theory, this
parameter is the so-called Coulomb pseudopotential \cite{16}, which
is in no way related to any measurable parameter of the system, i.e.
is practically "bare". On the other hand, in the ATS, the adjustable
parameter should be on the order of average phonon frequency, and
numerical calculation results do confirm this theoretical prediction
\cite{15}. It is quite obvious that, had this study appeared in
1960s, physics of superconductivity could have an entirely different
history. However, what with the enormous bulk of experimental data
interpreted since the first paper on BCS theory appeared \cite{17},
rejecting the results of a gigantic work done by several generations
of theorists constitutes a huge psychological problem. In this
situation, ATS partisans find themselves constrained to select one
experiment after another to compare their interpretations in both
theories in the hope of finding differences of interpretation that
would be significant from the point of view of selection of the
right theory. In this paper, the Josephson effect was selected as
such an experiment bearing in mind the key role it had played for
recognition of the BCS theory as the standard theory of
superconductivity.

\section{\label{sec:level1}Contact between two identical normal metals}
We start examining the problem by formulating the Hamiltonian of a
tunnel junction in the presence of an electromagnetic field of the
type that can only be taken into account using a gradient
transformation. An example of such system can be suggested in the
form of a metal ring with a gap as shown in Fig. 1. When magnetic
flux $\Phi$ passes through the ring, its effect can be totally taken
into account using a gradient transformation, the phase difference
$\varphi$ of the gradient transformation on both sides of the gap
being linearly dependent on $\Phi$ , i.e. $\varphi  = 2\pi \Phi
/\Phi _0$,where $\Phi _0  = hc/e$. In all subsequent calculations,
we take $\hbar  = 1$, as usual.

\begin{figure}
\includegraphics{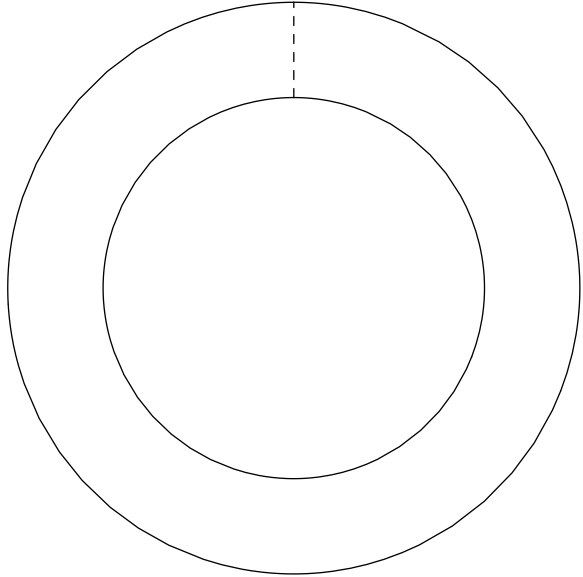}
\caption{\label{fig:epsart}Geometric setup of the system under
consideration. The dashed line represents the ring gap.}
\end{figure}

Consider a cubic lattice of dimensions $L_x  \times L_y  \times
L_z$. Let $ c_{\sigma {\bf g}}^ +$ and $c_{\sigma {\bf g}}$ be,
respectively, creation and annihilation operators of an electron
with a spin index $\sigma  =  \uparrow , \downarrow$ in the orbital
state, which forms the system conduction band, at a node ${\bf g}$;
then, for $ {\bf g}$:
\begin{equation}
\begin{array}{l}
{\bf g}_\alpha   = 1..N_\alpha, \\
N_\alpha   = L_\alpha  /a + 1,
\end{array}
\end{equation}
where $\alpha  = x,y,z$, and $a$ is the elementary cell size. Such a
system can provide a good description of the ring when $L_x \gg L_y
,L_z$. In this case $L_y$ and $L_z$ define the ring thickness, and $
L_x$ is equal to the ring perimeter.

Transformation into momentum representation defined with creation
operator $c_{\sigma {\bf p}}^ +$ and annihilation operator
$c_{\sigma {\bf p}}$ of an electron carrying a momentum ${\bf p}$ is
performed in a standard way:
\begin{equation}
\begin{array}{l}
c_{\sigma {\bf p}}^ +   = N_0^{ - 1/2} \sum\limits_{\bf g} {\exp
\left( {i\varphi \left( {a{\bf g}} \right)} \right)\sin \left( {{\bf
p}_x a{\bf g}_x } \right)\sin \left( {{\bf p}_y a{\bf g}_y }
\right)\sin \left( {{\bf p}_z a{\bf g}_z } \right)c_{\sigma {\bf
g}}^ +  },\\
 c_{\sigma {\bf p}}  = N_0^{ - 1/2} \sum\limits_{\bf g}
{\exp \left( { - i\varphi \left( {a{\bf g}} \right)} \right)\sin
\left( {{\bf p}_x a{\bf g}_x } \right)\sin \left( {{\bf p}_y a{\bf
g}_y } \right)\sin \left( {{\bf p}_z a{\bf g}_z } \right)c_{\sigma
{\bf g}} },
\end{array}
\end{equation}
where $ {\bf p}_\alpha   = \frac{{\pi {\bf \bar p}_\alpha  }}
{{L_\alpha   + 2a}}$, ${\bf \bar p}_\alpha   = 1..N_\alpha$, $N_0 =
\prod\limits_\alpha  {N_{0\alpha } }$, $ N_{0\alpha }  =
\frac{{L_\alpha   + 2a}} {{2a}}$, and $\varphi \left( {a{\bf g}}
\right)$ is the phase of the gradient transformation. In this form,
operators $ c_{\sigma {\bf p}}^+$ and $ c_{\sigma {\bf p}}$
correspond to eigenstates of the system. The reverse transformation
into the node representation is defined as follows:
\begin{equation}
\begin{array}{l}
c_{\sigma {\bf g}}^ +   = N_0^{ - 1/2} \exp \left( { - i\varphi
\left( {a{\bf g}} \right)} \right)\sum\limits_{\bf p} {\sin \left(
{{\bf p}_x a{\bf g}_x } \right)\sin \left( {{\bf p}_y a{\bf g}_y }
\right)\sin \left( {{\bf p}_z a{\bf g}_z } \right)c_{\sigma {\bf
p}}^ +  },\\
c_{\sigma {\bf g}}  = N_0^{ - 1/2} \exp \left( {i\varphi \left(
{a{\bf g}} \right)} \right)\sum\limits_{\bf p} {\sin \left( {{\bf
p}_x a{\bf g}_x } \right)\sin \left( {{\bf p}_y a{\bf g}_y }
\right)\sin \left( {{\bf p}_z a{\bf g}_z } \right)c_{\sigma {\bf p}}
}.
\end{array}
\end{equation}

Up to now, we considered the gap to be an ideal electric isolator.
Now, take into account the possibility of electron transfer between
the adjacent nodes on different sides of the gaps; to do so, we
introduce the tunneling Hamiltonian $\hat H_t$:
\begin{equation}
\hat H_t  =  - \gamma \sum\limits_\sigma  {\sum\limits_{{\bf g}_y
,{\bf g}_z } {c_{\sigma \left\{ {1,{\bf g}_y ,{\bf g}_z } \right\}}^
+  c_{\sigma\left\{ {N_x ,{\bf g}_y ,{\bf g}_z } \right\}} } }  +
H.c.
\end{equation}
Here, apparently, $\gamma$ is the hybridization potential, which we
take to be a real value measured in energy units. One can easily
obtain an expression for $\hat H_t$ in momentum representation:
\begin{equation}
\hat H_t  = \gamma \operatorname{e} ^{i\varphi } N_{0x}^{ - 1}
\sum\limits_\sigma  {\sum\limits_{{\bf p},{\bf k}} {\left( { - 1}
\right)^{{\bf \bar k}_x } \delta _{{\bf \bar k}_y }^{{\bf \bar p}_y
} \delta _{{\bf \bar k}_z }^{{\bf \bar p}_z } \sin \left( {{\bf p}_x
a} \right)\sin \left( {{\bf k}_x a} \right)c_{\sigma {\bf p}}^ +
c_{\sigma {\bf k}} } }  + H.c.
\end{equation}
where $\varphi$ is the phase difference between the sides of the
contact, i.e. $ \varphi  = \varphi \left( {aN_x ,a{\bf g}_y ,a{\bf
g}_z } \right) - \varphi \left( {a,a{\bf g}_y ,a{\bf g}_z }
\right)$, and $ \delta _{{\bf \bar k}_\alpha  }^{{\bf \bar p}_\alpha
}$ is the Kronecker delta. This expression can be transformed so as
to make the presence of the Josephson effect evident even in a
normal metal, although with a certain reserve:
\begin{equation}
\hat H_t  = N_{0x}^{ - 1} \sum\limits_\sigma  {\sum\limits_{{\bf
p},{\bf k}} {\left[ {\left( { - 1} \right)^{{\bf \bar k}_x }
\operatorname{e} ^{i\varphi }  + \left( { - 1} \right)^{{\bf \bar
p}_x } \operatorname{e} ^{ - i\varphi } } \right]\lambda \left(
{{\bf p},{\bf k}} \right)c_{\sigma {\bf p}}^ +  c_{\sigma {\bf k}} }
},
\end{equation}
where $\lambda \left( {{\bf p},{\bf k}} \right) = \gamma \delta
_{{\bf \bar k}_y }^{{\bf \bar p}_y } \delta _{{\bf \bar k}_z }^{{\bf
\bar p}_z } \sin \left( {{\bf p}_x a} \right)\sin \left( {{\bf k}_x
a} \right)$.

Change in the system energy  $\delta E^{\left( 1 \right)} \left(
\varphi  \right)$ is in the first order of perturbation theory over
$\lambda$ defined by the expression:
\begin{equation}
\delta E^{\left( 1 \right)} \left( \varphi  \right) = 2N_{0x}^{ - 1}
\cos \left( \varphi  \right)\sum\limits_{\sigma ,{\bf p}} {\left( {
- 1} \right)^{{\bf \bar p}_x } \lambda \left( {{\bf p},{\bf p}}
\right)n_{\sigma {\bf p}} }.
\end{equation}

A simple analysis of this expression makes one conclude that, at
certain relations between $L_x$, $L_y$ and $L_z$ the minimum of the
energy system can be displaced to a position of $\varphi  = \pi$ ,
but the energy gain obtained in transition to this state tends to
zero as the system size increases, i.e. one deals with a so-called
quantum dimension effect \cite{18}. This effect can be used in
attempts of creating matrices featuring unusual magnetic properties.

As for the second order of perturbation theory, it yields the
following expression for the correction $\delta E^{\left( 2 \right)}
\left( \varphi  \right)$:
\begin{equation}
\delta E^{\left( 2 \right)} \left( \varphi  \right) =
\frac{1}{2}N_{0x}^{ - 2} \sum\limits_\sigma  {\sum\limits_{{\bf p}
\ne {\bf k}} {\left| {\left( { - 1} \right)^{{\bf \bar k}_x }
{\mathop{\rm e}\nolimits} ^{i\varphi }  + \left( { - 1}
\right)^{{\bf \bar p}_x } {\mathop{\rm e}\nolimits} ^{ - i\varphi }
} \right|^2 \lambda ^2 \left( {{\bf p},{\bf k}}
\right)\frac{{n_{\sigma {\bf p}}  - n_{\sigma {\bf k}}
}}{{\varepsilon _{\bf p}  - \varepsilon _{\bf k} }}} } .
\end{equation}

These corrections are related to the electron transition through the
junction, which is why there exists the following obvious relation
$\delta E\left( \varphi  \right) - \delta E\left( 0 \right) =
\int\limits_0^t {I\left( {t'} \right)V\left( {t'} \right)dt'}  = e^{
- 1} \int\limits_0^\varphi  {I\left( {\varphi '} \right)d\varphi
'}$, where $e$ is the electron charge and $V$ is the circuit e.m.f.
This can easily be seen considering the tunneling current operator,
which can be expressed as:
\begin{equation}
\hat I_t  =  - ie\gamma \sum\limits_\sigma  {\sum\limits_{{\bf g}_y
,{\bf g}_z } {c_{\sigma \left\{ {1,{\bf g}_y ,{\bf g}_z } \right\}}^
+  c_{\sigma \left\{ {N_x ,{\bf g}_y ,{\bf g}_z } \right\}} } }  +
H.c.
\end{equation}
or, in the momentum representation, as:
\begin{equation}
\hat I_t  =  - ieN_{0x}^{ - 1} \sum\limits_\sigma {\sum\limits_{{\bf
p},{\bf k}} {\left[ {\left( { - 1} \right)^{{\bf \bar k}_x }
{\mathop{\rm e}\nolimits} ^{ - i\varphi }  - \left( { - 1}
\right)^{{\bf \bar p}_x } {\mathop{\rm e}\nolimits} ^{i\varphi } }
\right]\lambda \left( {{\bf p},{\bf k}} \right)c_{\sigma {\bf k}}^ +
c_{\sigma {\bf p}} } }.
\end{equation}

Indeed, corrections of the first and second order over $\lambda$ to
a current averaged over the ground state are of the form of:
\begin{equation}
\begin{array}{l}
 I^{\left( 1 \right)} \left( \varphi  \right) =  - 2eN_{0x}^{ - 1} \sin \left( \varphi  \right)\sum\limits_{\sigma ,{\bf p}} {\left( { - 1} \right)^{{\bf \bar p}_x } \lambda \left( {{\bf p},{\bf p}} \right)n_{\sigma {\bf p}} } , \\
 I^{\left( 2 \right)} \left( \varphi  \right) =  - 2e\sin \left( {2\varphi } \right)N_{0x}^{ - 2} \sum\limits_\sigma  {\sum\limits_{{\bf p} \ne {\bf k}} {\left( { - 1} \right)^{{\bf \bar p}_x  - {\bf \bar k}_x } \lambda ^2 \left( {{\bf p},{\bf k}} \right)\frac{{n_{\sigma {\bf p}}  - n_{\sigma {\bf k}} }}{{\varepsilon _{\bf p}  - \varepsilon _{\bf k} }}} } , \\
 \end{array}
\end{equation}
where $\varepsilon _{\bf p}$ and $\varepsilon _{\bf k}$ are energies
of electrons carrying, respectively, momenta ${\bf p}$ and ${\bf
k}$. The obtained expressions confirm the validity of the above
relation, which links the system energy change to the current
running through the contact.

It becomes evident that the Josephson effect is present even in a
normal metal, although it is so small in macroscopic samples that
one actually deals with an oscillating series with terms of similar
magnitude.

Curiously enough, when only the second-order correction $\delta
E^{\left( 2 \right)} \left( \varphi  \right)$ is taken into account,
Josephson transition for a system of two normal metals is at zero
temperature a so-called $\pi$ -type junction since $\delta E^{\left(
2 \right)} \left( \varphi  \right)$ is at a minimum at $\varphi  =
\pi /2$ (in the BCS theory, the parameter $\varphi$ has historically
been defined to be equal to $ 2e\int\limits_0^t {V\left( {t'}
\right)dt'}$). Indeed, when the values of ${\bf \bar p}_y$ and ${\bf
\bar p}_z$ are fixed, transition from filled to unfilled states
takes place with a change in parity of the value of ${\bf \bar
p}_x$. That means that the energy-related denominator in the above
expression is at a minimum when $ \left( { - 1} \right)^{{\bf \bar
p}_x  - {\bf \bar k}_x }  =  - 1$, which brings about this
conclusion.

\section{\label{sec:level1}Contact between two identical superconductors in the ATS}

Now consider a ring made of a superconductor. In the ATS, a
superconductor differs from a normal metal in that its effective
electron-electron interaction is strong enough to enable production
of coupled-pair states which, however, have nothing in common with
Cooper pairs. Presence of coupled states brings about appearance of
a gap in the single-particle excitation spectrum. Therefore,
considering the dynamics of a superconductor in single-particle
approximation is meaningless. Indeed, the spectrum gap produces a
delta dependence of group velocity of electrons on momentum, which
makes it difficult to describe the interaction of the system with an
electromagnetic field.

To describe the system in two-particle approximation, it is
convenient to use a formalism related to a transition to new
electron creation and annihilation operators \cite{15},
respectively, $C^+$ and $C$. From the mathematical point of view,
these operators have much in common with those introduced by
Kadanoff and Martin in their dynamical generalization of the static
BCS approach \cite{19}. A significant difference between the
approaches is only revealed at the physical level, when correlators
introduced by Kadanoff and Martin are defined.

The new representation allows easily obtaining the ground-state wave
function, as well as spectra of single- and two-particle
excitations. Parent operators $c^+$ and $c$ are related to the
operators $C^+$ and $C$ by the following expressions:
\begin{equation}
\begin{array}{l}
 c_{\sigma {\bf p}}^ +   = C_{\sigma {\bf p}}^ +   + \sum\limits_\nu  {\sum\limits_{{\bf k},{\bf q}} {\theta _{\sigma ,\nu ,{\bf q}}^{*{\bf k} - {\bf q},{\bf p} + {\bf q}} C_{\sigma {\bf p} + {\bf q}}^ +  C_{\nu {\bf k} - {\bf q}}^ +  C_{\nu {\bf k}} } } , \\
 c_{\sigma {\bf p}}  = C_{\sigma {\bf p}}  + \sum\limits_\nu  {\sum\limits_{{\bf k},{\bf q}} {\theta _{\sigma ,\nu ,{\bf q}}^{{\bf k} - {\bf q},{\bf p} + {\bf q}} C_{\nu {\bf k}}^ +  C_{\nu {\bf k} - {\bf q}} C_{\sigma {\bf p} + {\bf q}} } } , \\
 \end{array}
 \label{eq:twelve}
\end{equation}
where
\begin{equation}
\begin{array}{l}
 \theta _{\sigma ,\sigma ,{\bf q}}^{{\bf p},{\bf k}}  = \frac{1}{2}\left( {\delta _{{\bf \bar k} - {\bf \bar p}}^{{\bf \bar q}}  - \delta _{{\bf \bar q}}^0 } \right) + \chi _{1,{\bf q}}^{{\bf p},{\bf k}} , \\
 \theta _{\sigma , - \sigma ,{\bf q}}^{{\bf p},{\bf k}}  =  - \delta _{{\bf \bar q}}^0  + \chi _{0,{\bf q}}^{{\bf p},{\bf k}}  + \chi _{1,{\bf q}}^{{\bf p},{\bf k}} , \\
 \end{array}
\end{equation}
$\delta _{\bf n}^{\bf m}$ is the 3D Kronecker delta, i.e. $\delta
_{\bf n}^{\bf m}  = \prod\limits_\alpha  {\delta _{{\bf n}_\alpha
}^{{\bf m}_\alpha  } }$, and $\chi _{0,{\bf q}}^{{\bf p},{\bf k}}$
and $\chi _{1,{\bf q}}^{{\bf p},{\bf k}}$ are solutions of
eigenvalue problems for two-particle states with spins 0 and 1,
respectively.

The following commutation relations are valid for $\chi _{0,{\bf
q}}^{{\bf p},{\bf k}}$ and $\chi _{1,{\bf q}}^{{\bf p},{\bf k}}$
\cite {20}:
\begin{equation}
\begin{array}{l}
\chi _{0,{\bf q}}^{{\bf p},{\bf k}}  = \chi _{0,{\bf k} - {\bf p} -
{\bf q}}^{{\bf p},{\bf k}} ,\chi _{0,{\bf q}}^{{\bf p},{\bf k}}  =
\chi _{0, - {\bf q}}^{{\bf k},{\bf p}} ,\\
\chi _{1,{\bf q}}^{{\bf p},{\bf k}}  =  - \chi _{1,{\bf k} - {\bf p}
- {\bf q}}^{{\bf p},{\bf k}} ,\chi _{1,{\bf q}}^{{\bf p},{\bf k}}  =
\chi _{1, - {\bf q}}^{{\bf k},{\bf p}} .
\end{array}
\end{equation}

As for normalizing relations, they have the following form:
\begin{equation}
\begin{array}{l}
\sum\limits_{\bf x} {\chi _{0,{\bf x}}^{{\bf p},{\bf k} + {\bf q}}
\chi _{0,{\bf q} - {\bf x}}^{{\bf k},{\bf p} + {\bf q}} }  =
\frac{1} {2}\left( {\delta _{{\bf \bar q}}^0  + \delta _{{\bf \bar
k}}^{{\bf \bar p}} } \right),\\
\sum\limits_{\bf x} {\chi _{1,{\bf x}}^{{\bf p},{\bf k} + {\bf q}}
\chi _{1,{\bf q} - {\bf x}}^{{\bf k},{\bf p} + {\bf q}} }  =
\frac{1} {2}\left( {\delta _{{\bf \bar q}}^0  - \delta _{{\bf \bar
k}}^{{\bf \bar p}} } \right).
\end{array}
\end{equation}
The Hamiltonian $\hat H_t$ can now be expressed via the new
operators $C^+$ and $C^+$:
\begin{eqnarray}
   \hat H_t  &=& N_{0x}^{ - 1} \sum\limits_{{\bf p},{\bf p'}} {\left[ {\left( { - 1} \right)^{{\bf \bar p'}_x } \operatorname{e} ^{i\varphi }  + \left( { - 1} \right)^{{\bf \bar p}_x } \operatorname{e} ^{ - i\varphi } } \right]\lambda \left( {{\bf p},{\bf p'}} \right)\sum\limits_\sigma  {C_{\sigma {\bf p}}^ +  C_{\sigma {\bf p'}} } }   \nonumber\\& &
    - N_{0x}^{ - 1} \sum\limits_{{\bf p},{\bf p'}} {\left[ {\left( { - 1} \right)^{{\bf \bar p'}_x } \operatorname{e} ^{i\varphi }  + \left( { - 1} \right)^{{\bf \bar p}_x } \operatorname{e} ^{ - i\varphi } } \right]\lambda \left( {{\bf p},{\bf p'}} \right)\sum\limits_{\sigma ,\nu ,{\bf k}} {C_{\sigma {\bf p}}^ +  C_{\nu {\bf k}}^ +  C_{\nu {\bf k}} C_{\sigma {\bf p'}} } }   \nonumber\\& &
    + N_{0x}^{ - 1} \sum\limits_{{\bf p},{\bf p'}} {\left[ {\left( { - 1} \right)^{{\bf \bar p'}_x } \operatorname{e} ^{i\varphi }  + \left( { - 1} \right)^{{\bf \bar p}_x } \operatorname{e} ^{ - i\varphi } } \right]\lambda \left( {{\bf p},{\bf p'}} \right)\sum\limits_{\sigma ,\nu ,{\bf k},{\bf q},{\bf q'}} {\chi _{1,{\bf q}}^{{\bf k} - {\bf q},{\bf p} + {\bf q}*} \chi _{1,{\bf q'}}^{{\bf k} - {\bf q'},{\bf p'} + {\bf q'}} C_{\sigma {\bf p} + {\bf q}}^ +  C_{\nu {\bf k} - {\bf q}}^ +  C_{\nu {\bf k} - {\bf q'}} C_{\sigma {\bf p'} + {\bf q'}} } }   \nonumber\\& &
    + N_{0x}^{ - 1} \sum\limits_{{\bf p},{\bf p'}} {\left[ {\left( { - 1} \right)^{{\bf \bar p'}_x } \operatorname{e} ^{i\varphi }  + \left( { - 1} \right)^{{\bf \bar p}_x } \operatorname{e} ^{ - i\varphi } } \right]\lambda \left( {{\bf p},{\bf p'}} \right)\sum\limits_{\sigma ,{\bf k},{\bf q},{\bf q'}} {\chi _{0,{\bf q}}^{{\bf k} - {\bf q},{\bf p} + {\bf q}*} \chi _{0,{\bf q'}}^{{\bf k} - {\bf q'},{\bf p'} + {\bf q'}} C_{\sigma {\bf p} + {\bf q}}^ +  C_{ - \sigma {\bf k} - {\bf q}}^ +  C_{ - \sigma {\bf k} - {\bf q'}} C_{\sigma {\bf p'} + {\bf q'}} } }.  \nonumber\\& &
\end{eqnarray}
This expression does not include terms having the structure of $C^ +
C^ +  C^ +  CCC$, since those should only be taken into account in
three-particle approximation, which involves introducing additional
terms into right hand side of~(\ref{eq:twelve}).

One obtains for uncoupled states:
\begin{equation}
\begin{array}{l}
\chi _{0,\mathbf{q}}^{\mathbf{p},\mathbf{k}}  \approx \frac{1}
{2}\left( {\delta _{\mathbf{\bar q}}^0  + \delta _{\mathbf{\bar k} -
\mathbf{\bar p}}^{\mathbf{\bar q}} } \right),\\
\chi _{1,\mathbf{q}}^{\mathbf{p},\mathbf{k}}  \approx \frac{1}
{2}\left( {\delta _{\mathbf{\bar q}}^0  - \delta _{\mathbf{\bar k} -
\mathbf{\bar p}}^{\mathbf{\bar q}} } \right).
\end{array}
\end{equation}
The non-strict equalities in these expressions are related to
corrections that are similar to Born approximation corrections in
the scattering theory, which we neglect.

There is every reason to assume \cite{15} that in the simplest
low-temperature superconductors, coupled states are only produced
for zero-spin states, either at a single node of the momentum
lattice or at adjacent nodes. Then, one may hereafter consider the
following relations to be satisfied:
\begin{equation}
\begin{array}{l}
\chi _{0,\mathbf{q}}^{\mathbf{p},\mathbf{k}}  = \frac{1} {2}\left(
{\delta _{\mathbf{\bar q}}^0  + \delta _{\mathbf{\bar k} -
\mathbf{\bar p}}^{\mathbf{\bar q}} } \right) +
h_\mathbf{q}^{\mathbf{p},\mathbf{k}},\\
\chi _{1,\mathbf{q}}^{\mathbf{p},\mathbf{k}}  = \frac{1} {2}\left(
{\delta _{\mathbf{\bar q}}^0  - \delta _{\mathbf{\bar k} -
\mathbf{\bar p}}^{\mathbf{\bar q}} } \right),
\end{array}
\end{equation}
$h_\mathbf{q}^{\mathbf{p},\mathbf{k}}$ is only non-zero when $\left|
{\mathbf{p}_\alpha   - \mathbf{k}_\alpha  } \right| < 2$, and it
only is in these cases that:
\begin{equation}
h_\mathbf{q}^{\mathbf{p},\mathbf{k}}  =  - \frac{1} {2}\left(
{\delta _{\mathbf{\bar q}}^0  + \delta _{\mathbf{\bar k} -
\mathbf{\bar p}}^{\mathbf{\bar q}} } \right) + \Omega ^{ - 1/2} \psi
^{\mathbf{p},\mathbf{k}} \left( \mathbf{q} \right),
\end{equation}
where $\psi ^{\mathbf{p},\mathbf{k}} \left( \mathbf{q} \right)$ is
the coupled-state wave function, and $\Omega  = \pi ^{ - 3}
\prod\limits_\alpha  {L_\alpha  }$. If one believes the effective
potential of the electron-electron interaction to be weakly
dependent on electron momenta and spherically symmetric, the above
normalization relations then yield:
\begin{equation}
\psi ^{\mathbf{p},\mathbf{p}} \left( \mathbf{q} \right) \approx \psi
_0 \left( q \right), \label{eq:twenty}
\end{equation}
and with  $\mathbf{p} \ne \mathbf{k}$,
\begin{equation}
\psi ^{\mathbf{p},\mathbf{k}} \left( \mathbf{q} \right) \approx \psi
_0 \left( q \right)/\sqrt 2, \label{eq:twentyone}
\end{equation}
while $\int {\left| {\psi _0 \left( q \right)} \right|^2
d\mathbf{q}} = 1$. Difference between the expressions
~(\ref{eq:twenty}) and ~(\ref{eq:twentyone}) reflects the fact that
two electrons of opposite spins located at the same node form a
zero-spin state, and the probability of such a state to be formed at
different nodes is equal to 1/2.

Expanding of the Hamiltonian $\hat H_t$ in powers of
$h_\mathbf{q}^{\mathbf{p},\mathbf{k}}$ reveals that it consists of
four terms. The first term $\hat H_t^{\left( 1 \right)}$ simply
reproduces the Hamiltonian of a normal metal, symbols $C$
substituting $c$. Of course, the terms $\hat H_t^{\left( 1 \right)}$
yield an oscillating series with terms of similar magnitude and do
not contribute to Josephson current for macroscopic systems. The
second term $\hat H_t^{\left( 2 \right)}$ is related to Kronecker
deltas contained in $h_\mathbf{q}^{\mathbf{p},\mathbf{k}}$ and has
the form of:
\begin{eqnarray}
\hat H_t^{\left( 2 \right)}  &=&  - \frac{1} {4}N_{0x}^{ - 1}
\sum\limits_{\left| {\mathbf{p}_\alpha   - \mathbf{k}_\alpha  }
\right| < 2} {\sum\limits_{\mathbf{p'}} {\left[ {\left( { - 1}
\right)^{\mathbf{\bar p'}_x } \operatorname{e} ^{i\varphi }  +
\left( { - 1} \right)^{\mathbf{\bar p}_x } \operatorname{e} ^{ -
i\varphi } } \right]\lambda \left( {\mathbf{p},\mathbf{p'}}
\right)\sum\limits_\sigma  {C_{\sigma \mathbf{p}}^ +  C_{ - \sigma
\mathbf{k}}^ +  C_{ - \sigma \mathbf{k}} C_{\sigma \mathbf{p'}} } }
}\nonumber\\& &
 - \frac{1}
{4}N_{0x}^{ - 1} \sum\limits_{\left| {\mathbf{p}_\alpha   -
\mathbf{k}_\alpha  } \right| < 2} {\sum\limits_{\mathbf{p'}} {\left[
{\left( { - 1} \right)^{\mathbf{\bar p'}_x } \operatorname{e}
^{i\varphi }  + \left( { - 1} \right)^{\mathbf{\bar p}_x }
\operatorname{e} ^{ - i\varphi } } \right]\lambda \left(
{\mathbf{p},\mathbf{p'}} \right)\sum\limits_\sigma  {C_{\sigma
\mathbf{k}}^ +  C_{ - \sigma \mathbf{p}}^ +  C_{ - \sigma
\mathbf{k}} C_{\sigma \mathbf{p'}} } } }\nonumber\\& &
 - \frac{1}
{4}N_{0x}^{ - 1} \sum\limits_\mathbf{p} {\sum\limits_{\left|
{\mathbf{p'}_\alpha   - \mathbf{k'}_\alpha  } \right| < 2} {\left[
{\left( { - 1} \right)^{\mathbf{\bar p'}_x } \operatorname{e}
^{i\varphi }  + \left( { - 1} \right)^{\mathbf{\bar p}_x }
\operatorname{e} ^{ - i\varphi } } \right]\lambda \left(
{\mathbf{p},\mathbf{p'}} \right)\sum\limits_\sigma  {C_{\sigma
\mathbf{p}}^ +  C_{ - \sigma \mathbf{k'}}^ +  C_{ - \sigma
\mathbf{k'}} C_{\sigma \mathbf{p'}} } } }\nonumber\\& &
 - \frac{1}
{4}N_{0x}^{ - 1} \sum\limits_\mathbf{p} {\sum\limits_{\left|
{\mathbf{p'}_\alpha   - \mathbf{k'}_\alpha  } \right| < 2} {\left[
{\left( { - 1} \right)^{\mathbf{\bar p'}_x } \operatorname{e}
^{i\varphi }  + \left( { - 1} \right)^{\mathbf{\bar p}_x }
\operatorname{e} ^{ - i\varphi } } \right]\lambda \left(
{\mathbf{p},\mathbf{p'}} \right)\sum\limits_\sigma  {C_{\sigma
\mathbf{p}}^ +  C_{ - \sigma \mathbf{k'}}^ +  C_{ - \sigma
\mathbf{p'}} C_{\sigma \mathbf{k'}} } } }.\nonumber\\& &
\end{eqnarray}
Obviously, $\hat H_t^{\left( 2 \right)}$ also does not contribute to
Josephson current in macroscopic systems. The Hamiltonian $\hat
H_t^{\left( 3 \right)}  \sim \psi ,\psi ^*$ describes transition
between coupled and uncoupled states, and will not be reproduced
here due to its cumbersomeness. It does not contribute to Josephson
current either.

Now the fourth term  $\hat H_t^{\left( 4 \right)}  \sim \psi ^*
\psi$ describes a transition that is accompanied by production of a
coupled electron pair above the upper limit of the gap, and
production of a coupled hole pair below the lower limit of the
superconductivity gap of the electron system:
\begin{equation}
\begin{array}{l}
 \hat H_t^{\left( 4 \right)}  = N_{0x}^{ - 1} \Omega ^{ - 1}  \\
  \times \sum\limits_{{\bf p},{\bf k},{\bf p'},{\bf k'}} {\sum\limits_{\sigma ,{\bf q}} {\left( { - 1} \right)^{{\bf \bar k}_x  + {\bf \bar q}_x } \left[ \begin{array}{l}
 \left( { - 1} \right)^{{\bf \bar p'}_x  + {\bf \bar k'}_x } {\mathop{\rm e}\nolimits} ^{i\varphi }  \\
  + \left( { - 1} \right)^{{\bf \bar p}_x  + {\bf \bar k}_x } {\mathop{\rm e}\nolimits} ^{ - i\varphi }  \\
 \end{array} \right]\lambda \left( \begin{array}{l}
 {\bf p} - {\bf q}, \\
 {\bf p'} + {\bf k'} - {\bf k} - {\bf q} \\
 \end{array} \right)\left[ \begin{array}{l}
 \psi ^{{\bf k},{\bf p}*} \left( {\bf q} \right) \times  \\
 \psi ^{{\bf k'},{\bf p'}} \left( {{\bf k} - {\bf k'} + {\bf q}} \right) \\
 \end{array} \right]C_{\sigma {\bf p}}^ +  C_{ - \sigma {\bf k}}^ +  C_{ - \sigma {\bf k'}} C_{\sigma {\bf p'}} } }  \\
 \end{array}
\end{equation}
while ${\left| {{\bf p}_\alpha   - {\bf k}_\alpha  } \right| < 2}$,
${\left| {{\bf p'}_\alpha   - {\bf k'}_\alpha  } \right| < 2}$.

The Hamiltonian $\hat H_t^{\left( 4 \right)}$ can contribute to the
Josephson current of a macroscopic system in the second perturbation
theory order over $\lambda$. In order to demonstrate this, let us
consider the transitions defined by this Hamiltonian more in detail,
while bearing in mind that the total pair momentum along the
directions $y$ and $z$ is an integral of motion. First, consider the
case when ${\bf \bar p}_y  = {\bf \bar k}_y  = {\bf \bar p'}_y  =
{\bf \bar k'}_y$ and ${\bf \bar p}_z = {\bf \bar k}_z  = {\bf \bar
p'}_z  = {\bf \bar k'}_z$, respectively, i.e. the pair momenta are
located at the same node in the $yz$ plane of the momentum space.
Two coupled states with odd total quantum number (QN) $2{\bf \bar
p'}_x + 1$ can be produced: $C_{ \uparrow ,\left\{ {{\bf \bar p'}_x
,{\bf \bar p'}_y ,{\bf \bar p'}_z } \right\}}^ +  C_{ \downarrow
,{\bf \bar p'}_x  + 1,{\bf \bar p'}_y ,{\bf \bar p'}_z }^ +  \left|
0 \right\rangle$ and $C_{ \uparrow ,\left\{ {{\bf \bar p'}_x  +
1,{\bf \bar p'}_y ,{\bf \bar p'}_z } \right\}}^ +  C_{ \downarrow
,{\bf \bar p'}_x ,{\bf \bar p'}_y ,{\bf \bar p'}_z }^ +  \left| 0
\right\rangle$. In the same way, two coupled states can be produced
with odd total QN $2{\bf \bar p}_x  + 1$: $C_{ \uparrow ,\left\{
{{\bf \bar p}_x ,{\bf \bar p}_y ,{\bf \bar p}_z } \right\}}^ +  C_{
\downarrow ,\left\{ {{\bf \bar p}_x  + 1,{\bf \bar p}_y ,{\bf \bar
p}_z } \right\}}^ +  \left| 0 \right\rangle$ and $C_{ \uparrow
,\left\{ {{\bf \bar p}_x  + 1,{\bf \bar p}_y ,{\bf \bar p}_z }
\right\}}^ +  C_{ \downarrow ,{\bf \bar p}_x ,{\bf \bar p}_y ,{\bf
\bar p}_z }^ +  \left| 0 \right\rangle$. Therefore, there exist 4
parity-conserving transitions $2{\bf \bar p'}_x  + 1 \to 2{\bf \bar
p}_x  + 1$, and the squared modulus of the corresponding matrix
element has a factor of 1/4 related to $\psi ^{{\bf p},{\bf k}}
\left( {\bf q} \right)$ normalization at ${\bf p} \ne {\bf k}$. In
this way, the produced integral factor for energy change in the
second order of perturbation theory over $\lambda$ in $2{\bf \bar
p'}_x  + 1 \to 2{\bf \bar p}_x  + 1$ transitions turns out to be
equal to 1.

For each of the total QNs $2{\bf \bar p'}_x$ and $2{\bf \bar p}_x$,
there exists one coupled state: $C_{ \uparrow ,\left\{ {{\bf \bar
p'}_x ,{\bf \bar p'}_y ,{\bf \bar p'}_z } \right\}}^ + C_{
\downarrow ,\left\{ {{\bf \bar p'}_x ,{\bf \bar p'}_y ,{\bf \bar
p'}} \right\}}^ +  \left| 0 \right\rangle$ and $C_{ \uparrow
,\left\{ {{\bf \bar p}_x ,{\bf \bar p}_y ,{\bf \bar p}_z }
\right\}}^ +  C_{ \downarrow ,\left\{ {{\bf \bar p}_x ,{\bf \bar
p}_y ,{\bf \bar p}_z } \right\}}^ +  \left| 0 \right\rangle$,
respectively. Therefore, there exists one transition $2{\bf \bar
p'}_x  \to 2{\bf \bar p}_x$; taking into account the normalization
of the function $\psi ^{{\bf p},{\bf p}} \left( {\bf q} \right)$, it
also contributes to the said integral factor, its contribution also
being equal to 1.

Parity-violating transitions, symbolically expressed as $2{\bf \bar
p'}_x  + 1 \to 2{\bf \bar p}_x$ and $2{\bf \bar p'}_x  \to 2{\bf
\bar p}_x  + 1$,  are four; their squared matrix elements have
factors of 1/2, and the contributions of parity-violating
transitions to the integral factor is equal to -2. Thus,
contributions of parity-conserving and parity-violating transitions
to the integral factor cancel each other when the pair momenta are
located at the same node of the $yz$ plane in the momentum space.

Consider now the case where the momenta of the coupled pairs are
located at neighboring nodes of the $yz$ plane in the momentum
space. These are best considered together as in this case, there is
no difference in normalizations of coupled wave functions $\psi
^{\mathbf{p},\mathbf{k}} \left( \mathbf{q} \right)$, because always,
${\bf p} \ne {\bf k}$. Assume that there are $m$ initial and final
states with odd values of the QN under consideration, and,
respectively, $n$ initial and final states with even QN values in
the corresponding unit intervals of the momentum space. Then, the
number of parity-conserving transitions is $m^2  + n^2$, and that of
parity-violating transitions, $2mn$. Therefore, if $m \ne n$ then
the number of parity-conserving transitions would be larger than
that of parity-violating transitions. On the other hand, evidently,
$m = 2n$ (which is confirmed by the above detailed analysis of the
particular case where the couple momenta are located at the same
node of the $yz$ plane in the momentum space). Thus, the
contribution to energy change is larger from parity-conserving
transitions than from parity-violating transitions in the case where
the momenta of the coupled pairs are located in the neighboring
nodes of the $yz$ plane in the momentum space. Consequently, an
uncompensated Josephson current is produced, and the ground state of
the system is a state with $\varphi  = 0$, i.e. it is a normal
Josephson junction. Besides, the presence of the term $\psi ^{{\bf
k},{\bf p}*} \left( {\bf q} \right) \times \psi ^{{\bf k'},{\bf p'}}
\left( {{\bf k} - {\bf k'} + {\bf q}} \right)$ in the right-hand
part of (23) allows one to conclude that the Josephson current
density tends to zero as the pair coupling energy $E_b$ , linearly
related to the energy gap  as $\Delta  = 7E_b$ \cite{13}, decreases,
which agrees with the experimental data too.

\section{\label{sec:level1}Josephson effect in the BCS theory}

Consider a superconductor at zero temperature from the point of view
of the BCS theory. In this system, the ground state $\left| {BCS}
\right\rangle$ satisfies the following relation:
\begin{equation}
b_{\sigma {\bf k}} \left| {BCS} \right\rangle  = 0,
\end{equation}
where  $b_{\sigma {\bf k}}$ are Bogoliubov quasi-particle
annihilation operators related to electron ñcreation and
annihilation operators in the following way (taking  $\sigma  = \pm
1/2$):
\begin{equation}
\begin{array}{l}
b_{\sigma {\bf k}}^ +   = u_{0{\bf k}} c_{\sigma {\bf k}}^ +   -
2\sigma v_{0{\bf k}} c_{ - \sigma {\bf k}},\\
b_{\sigma {\bf k}}  = u_{0{\bf k}} c_{\sigma {\bf k}}  - 2\sigma
v_{0{\bf k}} c_{ - \sigma {\bf k}}^ +.
\end{array}
\end{equation}
We assume here that Bogoliubov transformation parameters $u_{0{\bf
k}}$ and $v_{0{\bf k}}$ are real values, i.e. in terms of the BCS
theory, the superconductor wave function phase is zero. Take now a
unitary transformation of the basis set of wave functions produced
using the $\left| {BCS} \right\rangle$ state and Bogoliubov
quasi-particle operators with the operator $\exp \left( {i\varphi
\hat N} \right)$, where $\hat N$ is the operator of total number of
electrons in the system ($\hat N = \sum\limits_{\sigma ,{\bf k}}
{c_{\sigma {\bf k}}^ +  c_{\sigma {\bf k}} }$). Now, the ground
state of the system is $\left| {BCS'} \right\rangle  = \exp \left(
{i\varphi \hat N} \right)\left| {BCS} \right\rangle$. It can easily
be seen after a transition to new Bogoliubov quasi-particle
operators $b_{\sigma {\bf k}}'^{+}$ and $b'_{\sigma {\bf k}}$:
\begin{equation}
\begin{array}{l}
 b_{\sigma {\bf k}}'^{+}   = u_{\bf k} c_{\sigma {\bf k}}^ +   - 2\sigma v_{\bf k} c_{ - \sigma {\bf k}},  \\
 b'_{\sigma {\bf k}}  = u_{\bf k}^* c_{\sigma {\bf k}}  - 2\sigma v_{\bf k}^* c_{ - \sigma {\bf k}}^ +  , \\
 \end{array}
\end{equation}
where
\begin{equation}
\begin{array}{l}
 u_{\bf k}  = u_{0{\bf k}} e^{i\varphi } , \\
 v_{\bf k}  = v_{0{\bf k}} e^{ - i\varphi } . \\
 \end{array}
\end{equation}
Obviously, the new ground state $\left| {BCS'} \right\rangle$ of the
superconductor satisfies the relation:
\begin{equation}
b'_{\sigma {\bf k}} \left| {BCS'} \right\rangle  = 0,
\end{equation}
since $b'_{\sigma {\bf k}}  = \exp \left( {i\varphi \hat N}
\right)b_{\sigma {\bf k}} \exp \left( { - i\varphi \hat N} \right)
$. The superconductor wave function in the BCS representation
acquires then a phase $2\varphi$, since  $u_{\bf k} v_{\bf k}^* \sim
\exp \left( {2i\varphi } \right)$. \emph{It is important to note
that the transformation  $\exp \left( {i\varphi \hat N} \right)$
does not correspond to any physical effect on the system}.

From the point of view of quantum physics, no transformation of the
basis set of wave functions can result in changes in observable
values, e.g. the system energy spectrum, as the matrix elements
corresponding to these values change as well. This, however, is not
quite correct in the BCS theory.

Indeed, consider two superconductors linked to each other by the
Hamiltonian $H_T$:
\begin{equation}
H_T  = \sum\limits_{\sigma ,{\bf k},{\bf q}} {T_{{\bf kq}} c_{\sigma
{\bf k}}^ +  c_{\sigma {\bf q}} }  + H.c.
\end{equation}
We assume the operators corresponding to momentum ${\bf k}$ to be
related to the electrons of the first superconductor, and those
corresponding to momentum ${\bf q}$ to be related to the electrons
of the second superconductor, respectively. Applying the
transformation $\exp \left( {i\varphi _1 \hat N_1 } \right)$ to the
basis wave function set of the first superconductor, and the
transformation $\exp \left( {i\varphi _2 \hat N_2 } \right)$
respectively, to the basis set of the second one, and expressing the
operators $c^ +$ and $c$ in terms of Bogoliubov operators
\begin{equation}
\begin{array}{l}
 c_{\sigma {\bf p}}^ +   = u_{\bf p}^* b_{\sigma {\bf p}}'^ +   + 2\sigma v_{\bf p} b'_{ - \sigma {\bf p}} , \\
 c_{\sigma {\bf p}}  = u_{\bf p} b'_{\sigma {\bf p}}  + 2\sigma v_{\bf p}^* b_{ - \sigma {\bf p}}'^ +  , \\
 \end{array}
\end{equation}
where ${\bf p} = {\bf k},{\bf q}$, allows calculating the
second-order correction $\delta E_2$ to the energy of the
two-superconductor system:
\begin{equation}
\delta E_2  =  - 2\sum\limits_{{\bf k},{\bf q}} {\left| {T_{{\bf
kq}} } \right|^2 \frac{{\left| {v_{\bf k} u_{\bf q}  + u_{\bf k}
v_{\bf q} } \right|^2 }}{{E_{\bf k}  + E_{\bf q} }}},
\label{eq:thirtyone}
\end{equation}
which is absolutely identical to the expression provided by Anderson
for binding energy of two superconductors at zero temperature (see
Eq.(3) in \cite{2}). This expression is the initial equation in
Josephson current calculations according to the BCS theory.

The right-hand part of~(\ref{eq:thirtyone}) depends on phase
difference $\varphi _2 - \varphi _1$ between the two
superconductors, which contradicts the principles of quantum
mechanics, since the observed values should not depend on the chosen
basis wave function set of the system. It should be noted once again
that the transformation $\exp \left( {i\varphi _j \hat N_j }
\right)$ ($j = 1,2$) of the basis set of two superconductors does
not correspond to any physical effect. Therefore, the expression for
Josephson current used in the BCS theory contradicts the principles
of quantum mechanics.

At this point, one might conclude examining the BCS theory
interpretation of the Josephson effect, leaving the experts on the
BSC theory to their own devices in their search for a way to resolve
the appearing paradox. Let us, however, try and assist our
colleagues in solving this tangled problem. In order to achieve this
goal we must return to the fundamentals of the BCS theory.

In the BCS theory the Hamiltonian of the system can be reduced to
the form $\hat H_0'  = \sum\limits_{\sigma ,{\bf p}} {E_{\bf p}
b_{\sigma {\bf p}}'^ +  b'_{\sigma {\bf p}} }$ with the use of a
variation procedure \cite{21}. This variation procedure is
interpreted by the BCS specialists as a generalization of the
Hartree-Fock approach for description of the single particle
spectrum of the system and enables the use of linear combinations of
wave functions with various numbers of electrons for the ground
state of the system. Of course, we could agree with such an
interpretation, would there be no such an important objection as the
following: the Hamiltonian $\hat H'_0$ does not preserve the total
number of the electrons in the system, i.e. one of the main system's
symmetries is broken. In such a situation the appearance of the
above-mentioned paradox in case of the use of the Hamiltonian $\hat
H'_0$ as the principle Hamiltonian in the perturbation theory is not
surprising. It is noteworthy to notice here that when building
various models for solids including those based on the Hartree-Fock
approximation physicists strictly observe the requirements imposed
on the symmetry of the system, that is why the interpretation of the
Hartree-Fock approximation  in the BCS theory cannot be called
felicitous.

Of course, the interpretation of the Hartree-Fock approximation
suggested by the authors of the BCS theory is not the only possible
one. Obviously, the principle of system description we had earlier
put forward \cite{15} also is a generalization of the Hartree-Fock
approximation, which, however, can be used to describe
\emph{multi}-particle spectra of the system. The advantage of our
version of the generalization is not reduced to the absence of the
above-mentioned paradox, this version also provides the possibility
of an arbitrarily precise description with extension of the
right-hand part of the~(\ref{eq:twelve}); a precise description of a
system containing $N_0$ electrons would only require $N_0$ terms to
be used in the right-hand part of~(\ref{eq:twelve}). One should
recall in this connection that all consistent physical theories,
with the apparent exception of the BCS theory, are required to
provide an arbitrarily precise description of the corresponding
systems. To our knowledge, at least, there are no studies where this
problem were consistently studied in the framework of this theory.

The above considerations suggest a way to resolving the paradox
related to the phase-difference dependence present in the right-hand
part of Eq.~(\ref{eq:thirtyone}). If our opponents do not for some
reasons agree with the suggested explanation of the paradox, we
would, of course, let them find their own way to eliminate this
dependence in order to keep the principles of quantum mechanics
inviolable. As for the conclusions of this study, the very fact of
existence of the paradox is what is of importance.

\section{\label{sec:level1}Conclusions.}

In my opinion, the process of acknowledgement of Josephson's and
Anderson's work results by the academic community is not quite
consistent. Indeed, if we are to acknowledge their interpretation of
the effect under discussion, then the interpretation of the basic
concepts of quantum mechanics should be changed, to supplement it
with a fundamentally new postulate of the dependence of measurement
results on the selected representation of the wave function of the
quantum object. In this case, it should also be admitted that modern
renderings of quantum physics undeservedly omit the references to
this contribution of the eminent scholars, which makes them,
strictly speaking, incorrect.

Assuming on the other hand that the Josephson effect might have a
different interpretation that does not require changing the bases of
quantum physics, such as for example put forward in the present
paper, one should reject a vast amount of interpretations of
experimental data produced in the framework of the BCS theory. It
should be reminded in this context that the problem of the simplest
BCS Hamiltonian cannot be solved using the quasi-spin operator
methods either \cite{22}, which is why all the results of the theory
have been obtained using the anomalous expectation values technique
or the equivalent Bogoliubov operator technique.

It appears that the academic community should either complete the
process of acknowledgement of the BCS theory or request the
physicists to create a different theory of superconductivity.

\begin{acknowledgments}
I am grateful to Prof. A. A. Rukhadze for a discussion of obtained
results, and to Dr. A. I. Golovashkin for his assistance in
preparation of the publication.
\end{acknowledgments}

\bibliography{Comparison}

\end{document}